\documentclass[12 pt]{article}

\begin{document}

\title{The Schwinger-DeWitt technique in Gauge-Gravity
Theories} \maketitle
\begin{center}
\author{Kanokkuan Chaicherdsakul}

\emph{Department of Physics, \\The University of Texas at Austin,
\\Austin, TX}

\end{center}

 Talk
presented at DPF (Division of Particle and Fields) $02$
conference, May $24-28,2002$ Williamburg, Virginia\footnote{The
link of this talk is at www.dpf2002.org/particle-astrophysics.cfm}

\begin{center} \textbf{Abstract}\end{center}
We construct the one-loop effective action in Yang-Mills and Pure
Quantum Gravity theories with heat kernel(or proper time method),
which maintains manifest covariance during and after quantization
(gauge and diffeomorphism invariance are always preserved).  In
this talk, we will basically focus on "What, How, and Why" we
prefer heat kernel than the standard Feynman diagram calculation
in momentum space at the one loop correction. The beta function of
Yang-Mills field in the fixed gravitational background can be more
simply obtained. The non-local term which cannot be easily
obtained in the expansion method are exactly computed in
Yang-Mills in the case of covariantly constant background field.
The local term is consistent with asymptotic expansion method or
any most standard method. The non-local terms give some physical
implication concerning non-perturbative problems such as
confinement and instabilities. The modification of this technique
to quantum gravity is discussed.

\section{Problems}

Progress in gauge-gravity theories depends on the development of
covariant formulation of quantum effective action. Background
field method is the elegant tool to construct effective action
while manifest invariant are always preserved during and after
quantization. However, most people follow the procedure on
computing Feynman diagrams in momentum space and have to regulate
each diagrams separately. This way makes the calculation quite
troublesome and even worse if we do not perform expansion over the
mean field. In addition, it is uncomfortable to keep track on
calculations once the gauge invariant breaks down due to gauge
fixing condition or mass cut off. Moreover,the well understood
divergent diagrams can only give the local part of effective
action. If we want to study the infrared properties of Yang-Mills,
we also need the non-local parts to give us some information what
and how much we can learn non-perturbative dynamics.
Unfortunately, the direct method are too complicated to carry out
the summation. For the quantum field in curved space and quantum
gravity, the non-local part should contain, for instance, particle
creation effects, vacuum polarization, and Hawking radiation in
black holes. Using graviton propagator in momentum space brings
trouble and calculation can be easily lost.\\
The aim of this paper is to overcome such problems above (or at
least part of it) with an alternative approach called heat kernel
or Schwinger-DeWitt technique to construct the invariant formalism
of effective action. Heat kernel was first formulated by
Schwinger[1] long times ago and then it was later applied in
geometrical language by DeWitt[2,3]. Vilkovisky and Barvinsky
generalized this technique to nonminimal but causal operators of
any order[4]. Recently, Barvinsky and Mukanov[5] derive the
non-perturbative and non-local approach to the effective action in
space time $ n
> 2 $ by summation of proper time series on the heat kernel and
suggest the new method. Our extension is somewhat different. We
constructed the exact expression of effective action which the
non-local terms are included by modifying the QED Schwinger's
results to Yang-Mills in the case of covariantly constant
background field. The results give implication about confinement
and instabilities. Additionally, the generalization of this
technique to quantum gravity is briefly discussed.

\section{Yang-Millls in Curved Space}
Pure Yang-Mills field in the fixed gravitational background in 4
dimension is considered. The action is
\begin{center}
  $ S = - \frac{1}{4 g_{y}^{2}}\int d^{4}x
  g^{\frac{1}{2}}F_{a\mu\nu}F^{a\mu\nu} \
$
\end{center}
 where $g_{y}$ is the Yang-Mills coupling constant.
 $g$ is
 defined as $|detg_{\mu\nu}|$. The indices $a$,$ b$, and $c$ are
 raised and lowered by Cartan Killing metric $\gamma_{ab}$
 \\ It is clear that this action is invariant under
 infinitesimal gauge transformation and also invariance under
 diffeomorphism (general coordinate transformation and reparameterization invariance)
 if the gravitational field is allowed to become dynamics in its
 own functional.\\
 Since the
 quadratic term in the mean field expansion is only taken to
 account at the one-loop level, the operator(called Jacobi Field operator)
 acting on the fluctuation field
 can be constructed by simply varying the classical field equation with
 imposing the differential supplementary condition.

\section{Schwinger's representation}
The Green's function has the formal representation defined as the
inverse of Jocobi field operator. Since the eigenvalues of
$\hat{F}$ are real, this can be replaced by a complex Laplace
transform or we just simply rotate by 90 degree in the complex
s-plane.
\begin{center}
$G = -\frac{1}{\hat{F}+i\epsilon} = i\int_{o}^{\infty}e^{i \hat{F}
s} ds$
\end{center}
and similarly for the ghost field operator
\begin{center}
$ G_{gh} = -\frac{1}{\hat{F_{gh}}+i\epsilon} =
i\int_{o}^{\infty}e^{i \hat{F_{gh}} s} ds$
\end{center}

 The good thing is that Schwinger's representation is valid when
  $\hat{F}$ is the Jacobi field
 operator of the {\it nonlinear} bosonic field and also valid when the
 background is {\it nonstationary} or in the
 arbitrary background[1,2].

\section{Unique Effective Action}
The effective action is formally defined as the Legendre transform
of the full connected one particle irreducible diagrams(1PI). In
usual formulation of gauge theories and quantum gravity, there are
two main problems. First, fixing the gauge of the quantized field
automatically fixes the gauge of the mean field and this leads to
non-gauge-invariant effective action. However, it is possible to
fix the gauge of the quantized field leaving the gauge of the mean
field arbitrary hence we obtain the gauge invariant effective
action. This is understood in background field gauge. The second
problem is that the gauge-invariant effective action depends
(parametrically) on the choice of background gauge conditions. The
theory is sensible only if the physical quantities do not depend
on the gauge choice. This problem was solved by DeWitt and
Vilkovisky[2,6]. That is why we call unique effective action which
is completely gauge independent\footnote{The gauge and ghost
independent in quantum gravity is discussed in Ref[10]}. To have
the inverse Green's function, the supplementary condition is
needed. After varying field equation and imposing supplementary
condition, we have the Jacobi field operator and ghost field
operator.
\begin{center}
$ \hat{^{a}_{\mu}F_{b}^{\nu}} =
\delta^{a}_{b}(\delta^{\mu}_{\nu}D_{\sigma}D^{\sigma}-R_{\mu}^{\nu})+
2f^{a}_{c b}F^{c \nu}_{\mu}$
\end{center}
\begin{center}
 $ \hat{F_{gh b}^{a}} =
\delta^{a}_{b}g^{\frac{1}{2}}D_{\mu}D^{\mu}$
\end{center}
One might ask why we do choose the supplementary condition in such
the way that the $ D^{\mu}D^{\nu} $ term is cancelled out from
the operator obtained from varying field equation. The reason is
that we want to preserve the manifest covariance and still have
the unique gauge invariant effective action. If the effective
action depends on the particular gauge chosen, it would be hard to
get the unambiguous and believable result.
\\
Let us recall the path integral quantization. For the one loop
quantum correction, we only need the
 Gaussian Integral or the quadratic term in the expansion.
\begin{center}
   $ W^{(1-loop)} = -i log (detG)^{\frac{1}{2}} + i log (detG_{gh}) = -\frac{i}{2} Trlog G + i Trlog G_{gh} $
 \end{center}
To construct effective action in term of heat kernel K, we apply
the Schwinger's representation.
\begin{center}
 $ W^{(1-loop)} = \frac{1}{2}Tr\int_{0}^{\infty}(e^{i \hat{F} s}- 2 e^{i\hat{F_{gh} } s})\frac{d s}{i s} $
\end{center}
Therefore, the $ w $ function is written as\footnote{The symbols
$Tr$ stands for the functional trace (space-time is included)
whereas $ tr $ means the trace over group indices.}
\begin{center}
$w(x)=\frac{1}{2}tr\int_{0}^{\infty}(K(x,x,s)-2K_{gh}(x,x,s))\frac{ds}{is}$
\end{center}
where
\begin{center}
$ W  =  \int d^{4}x w(x) $
\end{center}
and
\begin{center}
$ K(x,x',s)  =  e^{i\hat{F}s}\delta(x,x')$
 \end{center}
One can see that Schwinger's representation also gives the heat
equation.
 \begin{center}
 $ i \frac{\partial K}{\partial s} = - \hat{F} K $
\end{center}
and similar to the ghost kernel.

\section{Asymptotic expansion revisited}

The harder thing is to figure out what the kernel $K$ is in which
it can satisfy heat equations. If it is for the purpose of the
local part of effective action, proposing an ansatz kernel $K$ by
doing proper time expansion is good enough to investigate
ultraviolet divergence such as beta function and anomalies. It is
,therefore,

\begin{eqnarray*}
K(x,x',s) & = & i(4 \pi i
s)^{-2}D^{\frac{1}{2}}(x,x')e^{\frac{i\sigma}{2s}-im^{2}s}\Lambda(x,x',s)\\
  \Lambda(x,x',s) & = & \sum_{k=0}^{\infty}a_{k}(x,x')(is)^{k}\\
\end{eqnarray*}
(This is also similar for the ghost denoted by $K_{gh}$and $a_{k
gh}$. One can rotate by $ 90 $ degree to make sure that $a_{k}$
and $a_{gh k}$ are real) where
\begin{center}
  $ \Lambda(x,x,s=0) = 1$, $\sigma = \frac{1}{2}(x-x')^{2}$
\end{center}
and $ D^{\frac{1}{2}}(x,x')\equiv|det -\sigma_{;\mu\nu'}| $ is the
Van Vleck-Morette determinant. If one plug the kernel $K$ in $W$,
we see that the integral of effective action diverges at the lower
limit. How are we going to fix this? This asymptotic expansion
also admit the covariant regularization version. We choose
dimensional regularization for simplicity. The kernel $ K $ will
become $ K(x,x',s) = i(4 \pi i
s)^{-\frac{n}{2}}e^{\frac{i\sigma}{2s}-im^{2}s}\Lambda(x,x',s) $
and the integrand of effective action will diverge at the lower
limit for all positive value of n. The way out is to pretend that
the dimensionality n of space time is a complex number instead of
a positive integer and to define $w(x)$ by analytic continuation
to the region of the complex n-plane where integral converges to
the vicinity of actual physical dimension.
\\ Since only dimensionless quantities can be analytically
continued, one must multiply $ w(x) $ by $\mu^{-n}$. Inserting
kernel $K$ in $w(x)$ and integrating by part 4 times (for 4
dimensions). There will be coefficients up to $a_{2}$ needed to be
computed. After taking $m \rightarrow 0$, we obtain
\begin{center}
$ w = - \frac{\sqrt{g}}{32 \pi^{2}}(\frac{1}{n-4}-\frac{3}{4})tr(2
a_{2}-4 a_{2 gh}) + \Gamma_{ren}$
\end{center}
where $\Gamma_{ren}$ is the renormalized remainder after the local
terms have been subtracted.
\subsection{Coefficients and Coincidence limit}
As we see above, we need to know what the coefficient
$a_{2}(x,x')$ and $a_{gh 2}(x,x')$ precisely and take the limit $
x \rightarrow x' $ to determine the counter term. By inserting
kernel $ K $ and $ K_{gh} $in the heat equation, we get the
recursion relations.\footnote{The condense notation $ a_{;\mu} $
means covariant derivative with respect to $\mu $}
\begin{center}
$ \sigma_{;}^{\mu}a_{0;\mu} = 0$
\end{center}
\begin{center}
$ \sigma_{;}^{\mu} a_{k ; \mu} + k a_{k} =
\Delta^{-\frac{1}{2}}(\Delta^{\frac{1}{2}}a_{k-1})_{ ; \mu}^{\mu}
- (\delta^{a}_{b}R_{\nu}^{\mu}-2f^{a}_{b}F^{c\nu}_{\mu}) a_{k-1} $
\end{center}
\begin{center}
$ \sigma_{;}^{\mu} a_{gh k ; \mu} + k a_{gh k} =
\Delta^{-\frac{1}{2}}(\Delta^{\frac{1}{2}}a_{gh k-1})_{ ;
\mu}^{\mu} $
\end{center}
where $k = 1,2,...$ and $ \Delta(x,x')\equiv
g^{-\frac{1}{2}}D(x,x')g^{-\frac{1}{2}}$ \\
By using the commutation law of the Lie group, differentiating
these recursion relation,and taking the coincidence limit
$x\rightarrow x'$, we obtain
\begin{center} $tr[a_{gh 2}] = l
(\frac{1}{30}R_{;\mu}^{\mu}+\frac{1}{72}R^{2}-\frac{1}{180}R_{\mu\nu}R^{\mu\nu}+
\frac{1}{180}R_{\mu\nu\sigma\tau}R^{\mu\nu\sigma\tau})-
\frac{1}{12}F_{a\mu\nu}F^{a\mu\nu}$
\end{center}
\begin{center}
$tr[a_{2}] = l
(\frac{n-5}{30}R_{;\mu}^{\mu}+\frac{n-12}{72}R^{2}-\frac{n-90}{180}R_{\mu\nu}R^{\mu\nu}+
\frac{n-15}{180}R_{\mu\nu\sigma\tau}R^{\mu\nu\sigma\tau})+
\frac{24-n}{12}F_{a\mu\nu}F^{a\mu\nu}$
\end{center}
where $l$ is the dimension of Yang-Mills Lie group and n is the
space-time dimension.\\
The one-loop counter term is simply obtained by minimal
subtraction.
\begin{center} $ \delta(\frac{1}{g_{y}^{2}}) = - \frac{11}{24
\pi^{2}}\frac{1}{n-4}$
\end{center}

Notice that the renormalization group is irrelevant with the
curvature terms since only the right coefficients of
$F_{a\mu\nu}F^{a\mu\nu} $ term contribute beta function.
\section{(Exact) Solution in Yang-Mills}
Originally, heat kernel is understood as an expansion in the power
series of proper time variables. However, the effective action is
generically non-local and the calculation of non-local terms
requires the summation of the proper-time series. The direct
computation of the coefficients $ a_{k}$ is more difficult at
higher order of k . \emph{How are we going to fix this?} \\We are
more interested in the non-local effective action than the local
part (as already existed in most literature) to can interpret the
physical consequence. The original work was done in QED by
Schwinger[1] and in Yang-Mills and quantum gravity by DeWitt[2].
The more details are referred to these references. At present, we
briefly review what and how to get the results.
\\In fact, Schwinger have never used any kind of regularization to
renormalized QED.  He computed the kernel $ K(x,x',s)$ when $
F_{\mu\nu} $ is constant in Minkowski space time.
\begin{center}
$ A_{\mu} = - \frac{1}{2}F^{\mu\nu}x^{\nu}$
\end{center}
\begin{center}
\emph{What about in Yang-Mills?}
\end{center}
It would be less confusing if we choose the covariantly constant
Yang-Mills field $ F^{a}_{\mu\nu}$ in a particular direction of
the Lie algebra vector space. The simple Lie group in the adjoint
representation is assumed.
\begin{center}
$ F^{a}_{\mu\nu} = \delta^{a}_{1} F_{\mu\nu}$
\end{center}
\begin{center}
 $ A^{a}_{\mu} =
-\frac{1}{2} \delta_{1}^{a} F_{\mu\nu} x^{\nu}$
\end{center}
where $ F_{\mu\nu}$ is now acting as the electro-magnetic like
(chromo-electromagnetic field).\\
Therefore the heat equation becomes
\begin{center}
$ (i\frac{\partial}{\partial s} + (\partial_{\mu}+
f_{1}A^{1}_{\mu})(\partial^{\mu}+ f_{1}A^{1 \mu}))K_{gh} = 0 $
\end{center}
\begin{center}
 $
(i\frac{\partial}{\partial s} + (\partial_{\mu}+ f_{1}\otimes
1_{4}A^{1}_{\mu})(\partial^{\mu}+ f_{1}\otimes 1_{4}A^{1 \mu}) + 2
f_{1}\otimes \eta_{M}F\eta_{M}^{-1})K_{gh} = 0 $
\end{center}
where $ F \equiv F^{\mu}_{\nu}$, $ \eta_{M}\equiv \eta_{\mu\nu} $,
and $ f_{1}$ is the anti-symmetric matrix in Lie Algebra. Recall
\begin{center}
 $ w(x) = \frac{1}{2}\int_{0}^{\infty}tr (K- 2
 K_{gh})\frac{ds}{is} $
\end{center}
By diagolizing the matrices of heat equations, we have the generic
expression for the effective action.
\begin{center}
$ w(x) = \frac{i}{2} \int_{0}^{\infty}(4 \pi i s)^{-2}[2\rho +
det(\frac{F s}{sinh Fs})^{\frac{1}{2}}(\frac{1}{2}tr e^{2i
G_{1}\otimes\eta_{M}F\eta_{M}^{-1}s}-2)]\frac{ds}{is}$
\end{center}
where $ G_{1} = \left(%
\begin{array}{cc}
  0 & 1 \\
  -1 & 0 \\
\end{array}%
\right)$.\\
 After computing the determinant, we obtain
\begin{center}
$ w(x) = i \int_{0}^{\infty}(4 \pi i s)^{-2}[\rho - \frac{\chi
s^{2}}{Im cosks}(2 |cosks|^{2}-1)]\frac{ds}{is}$
\end{center}
where $ k = \sqrt{2(\alpha + i\chi)}$
\begin{center}
$\alpha  =  \frac{1}{4}F_{\mu\nu}F^{\mu\nu}  =
-\frac{1}{2}(\textbf{E}^{2}-\textbf{H}^{2})$
\end{center}

\begin{center}
$\chi =  \frac{1}{4}F_{\mu\nu}^{*}F^{\mu\nu}  =  \textbf{E} \cdot
\textbf{H} $
\end{center}
and $ \rho $ is the rank of Lie algebra.
 We will forget the dimensional regularization and instead
follow the Schwinger's way by subtracting the terms which cause
the integral divergence at $s \rightarrow 0$. The reason why we
can subtract is that, in any renormalizable theories, the coupling
constant or wave function renormalization can always be absorbed.
This subtraction procedure never make us losing consistency and
breaking gauge invariance. The quantities in the square brackets
in the integrand of effective action near $ s=0 $ has an expansion
in power s that begins $ (\rho+1) - \frac{11}{3}\alpha s^{2} +
O(s^{4})$ It is therefore our generic result is
\begin{center}
$ w = \frac{1}{16 \pi^{2}}\int_{0}^{\infty}[\frac{\chi}{s Im
cosks}(2|cosks|^{2}-1)+\frac{1}{s^{3}}]-\underbrace{\frac{11}{48
\pi^{2}}(\frac{1}{4}F_{\mu\nu}F^{\mu\nu})\int_{0}^{\infty}\frac{ds}{s}}$
\end{center}

* Notice that the last term is local and contribute the logarithmic
divergence part of effective action. As well known, its
coefficient naturally gives one-loop beta function consistent with
what is found in most literature or asymptotic expansion method
discussed earlier *
\\
The more interesting thing is how to extract the meaning of the
non-local effective action. Indeed, this corresponds to the
infinite summation of the proper time series. At present, we
restrict to the case of pure chromoelectric and pure
chromo-magnetic fields. After some algebra, the results are
\\
\emph{QCD results}:
\begin{center}
$ w^{YM}_{Mag.}  =   \frac{1}{16 \pi^{2}}
\int_{0}^{\infty}\frac{1}{s^{3}}(\frac{H s cos 2Hs}{sin Hs} + 1 -
\frac{11}{6} H^{2} s^{2}) ds$
\end{center}
\begin{center}
$ w^{YM}_{Elec.}  =  \frac{1}{16 \pi^{2}} \int_{0}^{\infty}
\frac{1}{s^{3}}( \frac{Es cosh 2Es}{sinh Es} + 1 + \frac{11}{6}
E^{2} s^{2}) ds $
\end{center}
Compared with
\\
\emph{QED results}[1,2]:
\begin{center}
$ w^{QED}_{Mag.}  =  \frac{1}{8 \pi^{2}}
\int_{0}^{\infty}\frac{1}{s^{3}}( H s cot Hs - 1 + \frac{1}{3}
H^{2} s^{2})e^{-im^{2}s} ds$
\end{center}
\begin{center}
$ w^{QED}_{Elec.}  =  \frac{1}{8 \pi^{2}}
\int_{0}^{\infty}\frac{1}{s^{3}}( E s coth Es - 1 - \frac{1}{3}
E^{2} s^{2})e^{-im^{2}s} ds$
\end{center}
Notice the signs of local terms between QED and QCD reverse due to
the asymptotic freedom of the Yang-Mills.
\section{Physical Interpretations}
\subsection{Asymptotic Freedom}
It is well known that the counter term obtained in the earlier
section implies
\begin{center}
$ \frac{1}{g_{y_{bare}}^{2}} =
\mu^{n-4}(\frac{1}{g_{y}^{2}}-\frac{11}{24\pi^{2}}\frac{1}{n-4})$
\end{center}
Instead of solving renormalization group equation, we can simply
differentiate this expression with respect to the energy scale
$\mu$. Hence we obtain the $\beta$ function.
\begin{center}
$ \beta = - \frac{11}{48 \pi^{2}}g_{y}^{3}$
\end{center}
where $ \beta $ is defined as $ \mu \frac{dg_{y}}{d\mu}$. The
running coupling constant is simply obtained.

\begin{center}
$ \frac{1}{g^{2}(\mu)} = \frac{1}{g_{0}^{2}} + \frac{11}{24
\pi^{2}} log\frac{\mu}{\mu_{0}}$
\end{center}
where $ g_{0}$ and $\mu_{0}$ are integration constants.
\\ As known, this is clearly indicated that the coupling
constant gets stronger at the high energy limit and weaker at the
low energy limit. In fact, this property is analogous to
non-linear sigma model on the Riemanian manifold whose target
space is the sphere $ S^{N}$. Suppose we couple the gluon field
with the fermionic fileds, it tends to destroy the effect.
\subsection{Confinement of pure electric Yang-Mills type}
In the case of covariantly constant chromo-electric field, there
is no way we can control the divergence. Note that this is
\emph{not} UV divergence as usual. The divergence stems from the
$cosh2Es $ term which is non-local. This result leads us to
conjecture that the theory is sensible(finite) only if the
background field of electric Yang-Mills type component are strong
enough to can satisfy the condition
\begin{center}
$ |\partial^{2}E| \gg E^{2}$
\end{center}
Suppose there is the regime in which the Yang-Mills field behave
like the Coulomb fields,the field will satisfy

\begin{center}
$ E \sim \frac{g_{y}^{2}}{r^{2}}$
\end{center}
But these two equations will be consistent only if $ g_{y}^{2} \ll
1$ and indicate that the field cannot behave like the Coulomb
filed in the strong coupling regime. The field must therefore
"crinkle". At least, this one-loop calculation gives some
implication about
confinement.\\
We later found that the model of confinement in chromo-electric
flux tube is studied in Ref[7]. It may be worth to consider
Yang-Mills in the strong background field to prove the conjecture
and study low energy QCD.

\subsection{Instabilities of pure magnetic Yang-Mills type}
If we compare the results of $w_{Mag}^{YM}$ and $w_{Mag}^{QED}$,
the integrand of $w_{Mag}^{QED}$ has the poles on the real axis.
This can be rotated to the negative imaginary axis without picking
up any residues, yielding the real valued QED effective action.\\
However, this is \emph{not} true in Yang-Mills. There are  the
infinite number of poles on the real axis. We cannot rotate to the
imaginary axis of the s-plane as in QED since we have no
exponential factor of mass term attached to it. The contribution
from the poles at $ s = \frac{n \pi}{H}, n = 1,2,3,...$ can be
done by running the integration contour just below the positive
real axis. The expression $w_{Mag}^{YM}$ will therefore become a
principle-value integral plus a sum of integrals over semicircles
around the poles. The residues at the poles are a sum of infinite
series. One has the question whether this series converges or not.
Fortunately,this is not too hard to can compute. We obtain the
finite series in the imaginary part. The result is
\begin{center}
$ w^{ren.}_{Mag.YM} = \frac{1}{16 \pi^{2}}P
\int_{0}^{\infty}(\frac{H cos 2Hs}{s^{2} sin Hs}+ \frac{1}{s^{3}}
- \frac{11}{6}\frac{H^{2}}{s}) ds + \frac{\pi i}{16 \pi^{2}}
\sum_{n=1}^{\infty} Res [...]  $

\end{center}
where $[...]$ stands for the integrand of the integral.
\begin{center}
$\sum_{n=1,3,...}Res [...]= -\frac{H^{2}}{\pi^{2}}\sum_{n=1,3,...}
\frac{1}{n^{2}}$
\end{center}
\begin{center}
$ \sum_{n=
2,4,...}Res[...]=\frac{H^{2}}{\pi^{2}}\sum_{n=2,4,..}\frac{1}{n^{2}}$
\end{center}
Hence

\begin{center}
$ Im w = - \frac{1}{192 \pi} H^{2}$
\end{center}
Notice that Nielsen and Olesen[8] found\footnote{The coefficient
factor of imaginary part is different but the real
 part(the logarithmic divergent term) is consistent.
  Note that we also have the non-local term contributed} $ Im W = - \frac{V
}{8\pi}H^{2} $\\
The reason why it contributes negative sign, unlike QED, is due to
the asymptotic freedom properties of the
 Yang-Mills fields.
 If $w$
-function is allowed to be imaginary, the vacuum state is unstable
and the true ground state is far from trivial. That means the
effective potential leads to the presence of negative mode
function.
\\We have studied the question of infrared instability of constant
chromomagnetic background on the series of paper of Ref[8]. There
are a lot of arguments.\\
The interpretations are summarized as following\\
1.It is implicitly Non-Perturbative problem. \\ 2. There is the
whole class of quantum runaway fluctuations, corresponding to mode
function with negative frequency at the
one-loop correction.\\
3.The vacuum is unstable. The true ground state cannot be found since the energy
density has developed an imaginary part.\\
4. The unstable mode corresponds to a \emph{tachyon} in 1+1 dimension[8]\\
5. To determine what mode exactly is stable and unstable,we have
to solve the eigenvalues equation and study its spectrum. \\
It is not yet perfectly clear how one precisely can stabilize this
chromo-magnetic type. The different authors give different
arguments. Flory suggested that the unstable mode can be
completely stabilized when one goes beyond one-loop calculation.
That means we need to keep the cubic and quartic terms in the
background field expansion. Avaraimi[8] concluded that it is
impossible to get a stable vacuum of chromo-magnetic type in
space-time of $n<5 $. What is the right way, we cannot tell at
this moment.

\section{One-loop Quantum Gravity}
The more interesting thing is to apply this technique to quantum
gravity. Let's start with pure quantum gravity action.
\begin{center}
$ S = 2\mu^{n-2}\int \sqrt{g}R d^{n}x$
\end{center}
where $\mu$ is the Planck mass and $R$ is the curvature scalar.\\
At present, we briefly review DeWitt's work[2]. The purpose is to
show that heat kernel can be applied in quantum gravity and is
less troublesome than the standard Feyman diagram calculation in momentum space. \\
We can apply the method described in the former sections to
construct the Jacobi field operators and ghost field operator.
After varying field equation and imposing supplementary condition,
we obtain
\begin{center}
$ _{\mu\nu}F^{\sigma\tau} =
\frac{1}{2}(\delta_{\mu}^{\sigma}\delta_{\nu}^{\tau}+\delta_{\mu}^{\tau}\delta_{\nu}^{\sigma})D_{\rho}D^{\rho}
-
\frac{1}{2}(\delta_{\mu}^{\sigma}\delta_{\nu}^{\tau}+\delta_{\mu}^{\tau}\delta_{\nu}^{\sigma}
- \frac{2}{n-2}g_{\mu\nu}g^{\sigma\tau})R -
\frac{2}{n-2}g_{\mu\nu}R^{\sigma\tau} - g^{\sigma\tau}R_{\mu\nu} +
 R_{\mu\nu}^{\sigma\tau} + R_{\mu\nu}^{\tau\sigma} +
\frac{1}{2}(\delta_{\mu}^{\sigma}R_{\nu}^{\tau}+\delta_{\mu}^{\tau}R_{\nu}^{\sigma}
+ \delta_{\nu}^{\sigma}R_{\mu}^{\tau} +
\delta_{\nu}^{\tau}R_{\mu}^{\sigma}) $
\end{center}
and
\begin{center}
$ F_{gh\nu}^{\mu} =
\sqrt{g}(\delta^{\mu}_{\nu}D_{\sigma}D^{\sigma}+ R^{\mu}_{\nu})$
\end{center}
In $4 $dimension, we need the $a_{2}$ coefficients at the
coincidence limit.(If we do in 2 dimension,for example, in
non-linear sigma model, we only need the $a_{1}$ coefficient) to
determine one loop counter term. The results are
\begin{center}
$tr[a_{2gh}] = \frac{n+5}{30} R_{;\mu}^{\mu} +
\frac{n+12}{72}R^{2}-\frac{n-90}{180}R_{\mu\nu}R^{\mu\nu}+
\frac{n-15}{180}R_{\mu\nu\sigma\tau}R^{\mu\nu\sigma\tau}$
\end{center}
and
\begin{center}
$ tr[a_{2}] = - \frac{n(2n-3)}{30}R_{;\mu}^{\mu} -
\frac{23n^{3}+145n^{2}-262n-144}{144(n-2)}R^{2} -
\frac{n^{3}+n^{2}+718n-720}{360(n-2)}R_{\mu\nu}R^{\mu\nu} +
\frac{n^{2}-29n+480}{360}R_{\mu\nu\sigma\tau}R^{\mu\nu\sigma\tau}$
\end{center}
Hence, we obtain one-loop $w$ function in quantum gravity
\begin{center}
$ w = - \frac{1}{16\pi^{2}(n-4)}\sqrt{g}(-
\frac{19}{15}R_{;\mu}^{\mu}- \frac{341}{36}R^{2}-
\frac{193}{90}R_{\mu\nu}R^{\mu\nu} +
\frac{19}{18}R_{\mu\nu\sigma\tau}R^{\mu\nu\sigma\tau}) -
\frac{1}{16\pi^{2}}\sqrt{g}(\frac{9}{20}R_{;\mu}^{\mu} +
\frac{459}{72}R^{2}-
\frac{103}{90}R_{\mu\nu}R^{\mu\nu}-\frac{31}{36}R_{\mu\nu\sigma\tau}R^{\mu\nu\sigma\tau})
-
\frac{i}{64\pi^{2}}\sqrt{g}\int_{0}^{\infty}ln(4i\pi\mu^{2})(\frac{\partial}{i\partial
s})^{3}[tr\Lambda(x,x,s)-2 tr\Lambda_{gh}(x,x,s)]ds $
\end{center}

Since quantum gravity is not perturbatively renormalizable, it has many difficulties. \\
$1.$ The effective action depend on the arbitrary constant $\mu$.
Each different choices of $\mu$ gives the different theories. By
setting $\mu \sim m_{Plank}$, it should give the roughly physical
interpretation at the Planck scale in order of magnitude
$10^{19}$ GeV. \\
$2.$ It contains the unknown coefficients of $a_{k}$ at higher
order of $k$. Up to now, partial solution exists. The coefficients
have been computed up to $a_{4}$[9] \\
$3.$ Suppose the renormalization is performed by minimal
subtraction then this last two lines should be added to the
classical action to get the effective action corrected to one-loop
order. As discussed earlier, we have the dependent auxiliary mass
problem.\\
Fortunately, there is the miracle in one-loop pure quantum
gravity. Because of the metric independence of the Euler-Poincare
characteristic, the counter term can be modified and obtained

\begin{center}
$ S_{c.t.
} = \frac{1}{16 \pi^{2}(n-4)}\int\sqrt{g}(-
\frac{429}{36}R^{2} + \frac{187}{90}R_{\mu\nu}R^{\mu\nu})$
\end{center}
Hence, it is (accidentally) one-loop finite on-shell.\\

It is interesting to see how one can construct (exact) effective
action in quantum gravity without repeating on iteration to
determine the coefficient $a_{k}$. One might need another
technique such as in-in formalism which might be a sensible
approach for solving quantum initial values problems. We will
leave this for investigation in the near future work.
\section{Conclusion and Outlook}
One should not jump to conclusions as to what is the true physics
at the strong coupling limit since what we have done is only based
on the one-loop approximation. Due to the fact that the sum over
all loops has physical significance, the physics can only be
decided by doing this sum. A straight forward approach certainly
leads to complexities beyond human power. However, it is possible
to generalize this technique to higher order loop. The calculation
is more complicated but not worse than the standard calculation in
momentum space. Additionally, other tricks on the correspondence
between gauge and gravity theories can be found
recently\footnote{The review is in Ref[10]}. The problems that
appear to be intractable on one side may stand a
chance of solution on the other side.  \\

In conclusion, we have developed the technique to construct
effective action in gauge and gravity theories. We found the
infrared instability in the massless chromomagnetic type which
stems from the imaginary part of effective action. The confinement
in the case of chromoelectric type is conjectured. The one loop
pure quantum gravity is finite. The connection between this
technique and closed-time path formalism in quantum gravity is
under investigation.

\section{Acknowledgement}
I thank the organizers at DPF02 conference for financial support
and the audiences for their interest and stimulating questions.
\section{References}
\begin{enumerate}
\item J. Schwinger, \emph{Phys. Rev. }\textbf{ 82 }, $ 664 (1951)$
\item B.S.DeWitt, \emph{The Global Approach to Quantum Field
Theory} ,Oxford University Press (2002) \item B. S.
DeWitt,\emph{Phys. Rept.}\textbf{19}, 295  (1975) \item
A.O.Barvinsky and G.A.Vilkovisky, \emph{Phys. Reports} \textbf{119
}, $  1-74 (1985)$ \item G.A. Vilkovisky, Nucl.Phys B \textbf{234}
$ 125 (1984)$
 \item A.O. Barvinsky and
V.F.Mukanov, hep-th/0203132 \item C. Flory \emph{Phys. Rev.
D}\textbf{29}, 722 (1984) \item N.K.Nielsen and P.Oleson,
\emph{Nucl. Phys. B}\textbf{144}, 378 (1978),H.B. Nielsen, M.
Ninomiya, \emph{Nucl. Phys.B} \textbf{156}, 1 (1979), C. Flory
e-print SLAC-PUB-3244 (1983),
Avramidi,\emph{J.Math.Phys.}\textbf{36}, 4 (1995) \item
 Avramidi,
\emph{Heat Kernel and Quantum Gravity}, Springer(2000)
 \item B.S.
DeWitt and C. Molina-Paris, \emph{Mod. Phys. Lett. A}
\textbf{13}2475 (1998) hep-th/9808163
 \item E. D'Hoker and D.Z. Freedman, \emph{TASI 2001 Lecture Notes}, hep-th/0201253
\end{enumerate}
\end{document}